\documentclass[letter,scriptaddress,twocolumn, prl,showkeys]{revtex4}
  \usepackage{amsmath}
  \usepackage{amssymb}
  \usepackage{makeidx}
  \usepackage{amsfonts}
  \usepackage[ansinew]{inputenc}
  \usepackage[usenames,dvipsnames]{pstricks}
  \usepackage{epsfig}
  \usepackage{pst-grad} 
  \usepackage{pst-plot} 
  \usepackage[colorlinks,hyperindex]{hyperref}
  \usepackage{lipsum}
  \usepackage{mathrsfs}
  \usepackage{mathtools}
  \hypersetup
  {
    colorlinks,%
    citecolor=black,%
    linkcolor=black,%
    urlcolor=black,%
  }

\makeindex

\begin{document}

\title {Quantum Teleportation with a Class of Non-Gaussian Entangled Resources}

\author{Soumyakanti Bose}
\email{soumyakanti@bose.res.in}
\affiliation{S. N. Bose National Centre for Basic Sciences \\ Block-JD, Sector-III, Salt Lake, Kolkata 700106 \\ India.\\
}
\author{M. Sanjay Kumar}
\email{sanjay@bose.res.in}
\affiliation{S. N. Bose National Centre for Basic Sciences \\ Block-JD, Sector-III, Salt Lake, Kolkata 700106 \\ India.\\
}


\begin{abstract}
Non-Gaussian entangled states of light have been found to improve the success of quantum teleportation. Earlier works in the literature focussed mainly on two-mode non-Gaussian states generated by de-Gaussification of two-mode squeezed vacuum states.
In the current work, we study quantum teleportation with a class of non-Gaussian entangled resource states that are generated at the output of a passive beam splitter (BS) with different input single mode non-Gaussian states. 
In particular, we consider input states that are generated under successive application of squeezing and photon addition/subtraction operations in various orders.
We focus on identifying what attributes of the resource states are necessary or sufficient for quantum teleportation (QT). 
To this end we first evaluate two attributes considered in the literature, viz. squeezed vacuum affinity (SVA) and EPR correlation.
While SVA is not non-zero for all two-mode resource states, EPR correlation is neither necessary nor sufficient of QT.
We consider yet another attribute, viz. two-mode quadrature squeezing as defined by Simon \emph{et. al.} [Phys. Rev. A \textbf{49}, 1567 (1994)].
Our numerical results on the de-Gaussified two-mode squeezed vacuum state as well as the BS generated non-Gaussian states lead us to the conclusion that two-mode quadrature squeezing is a {\it necessary condition} for QT, in general.
We further demonstrate the plausibility of this conclusion by giving an analytical proof that two-mode quadrature squeezing is a necessary condition for QT in the case of symmetric two-mode Gaussian resource states.
\end{abstract}


\maketitle
\section*{I. Introduction}

Quantum teleportation (QT) \cite{book_qit} is one of the most important information processing tasks that serves as a building block to several other protocols of quantum information technology \cite{qt_tp}.
It was first proposed by Bennett \emph{et. al.} \cite{TP_B} for qubits that could be realized in several systems such as atomic spin, polarization of light \emph{etc} \cite{book_qit}. 
Later, an experimentally realizable extension of the protocol to quantum optical systems was proposed by Braunstein and Kimble \cite{TP_BK}. 
Quantum teleportation with entangled optical resources, implementing Braunstein-Kimble (BK) protocol, has also been experimentally realized \cite{tp_exp1,tp_exp2,tp_exp3,tp_exp4,tp_exp5}. 

The most commonly used Gaussian entangled quantum optical resource in teleportation is the two-mode squeezed vacuum state (TMSV) which could be generated in parametric down conversion \cite{OPDC}.
However, certain de-Gaussification processes such as photon addition and subtraction  alongwith their coherent superposition, quantum catalysis \emph{etc.} have been found to improve the amount of entanglement as well as the success of teleportation compared to TMSV. \cite{ent_grangier,tp_illuminati,tp_yang,tp_lee,ent_benlloch,tp_wang, tp_agarwal,tp_zubairy}.

Dell'Anno \emph{et. al.} \cite{tp_illuminati} showed that optimized teleportation could be achieved by tuning entanglement, non-Gaussianity (NG) and squeezed vacuum affinity of the entangled resource state. 
Later developments \cite{tp_yang,tp_lee,tp_wang,tp_agarwal} have pointed to the possibility that Einstein-Podolosky-Rosen (EPR) correlation of the resource states could be a sufficient condition for QT. 
However, Lee \emph{et. al.} \cite{tp_lee} and Wang \emph{et. al.} \cite{tp_wang} have argued that EPR correlation is not always necessary for QT - for example, the symmetrically photon added TMSV yields QT even without EPR correlation.
Further, Hu \emph{et. al.} \cite{tp_zubairy} have addressed the question of whether there could be other aspects of the resource states, besides EPR correlation, that are crucial for QT. 
In this respect they considered the Hillery-Zubairy (HZ) correlation. However, they concluded that EPR correlation is a better witness of QT than HZ correlation, i.e., there exists resource states that yield QT that are not HZ correlated but are EPR correlated.
Clearly then, the question of what may be the \emph{necessary and sufficient condition} (s) for quantum teleportation is very much open.

It may be noted that all the non-Gaussian entangled states, in earlier works \cite{ent_grangier,tp_illuminati,tp_yang,tp_lee, ent_benlloch,tp_wang,tp_agarwal,tp_zubairy}, were generated by de-Gaussifying the TMSV. 
Another way to generate non-Gaussian entangled states is by using a passive BS with single mode nonclassical non-Gaussian states at one of the input ports. 
The BS output states are guaranteed to be entangled in view of the result on the necessary and sufficient condition for BS output entanglement \cite{nc_bsent}.
In our previous work \cite{bose_kumar}, we have studied various aspects of BS output entanglement with a class of input single mode non-Gaussian states, viz. the states that are generated under multiple nonclassicality inducing operations (MNIO) \cite{qs_mnc}.
In the present work, we explore such states in the context of quantum teleportation. 

The non-Gaussian resource states we consider here are generated under BS action with specific input states.
These input states are generated under successive application of various nonclassicality (NC)-inducing operations, viz., photon addition/subtraction and quadrature squeezing on the single mode vacuum.
The specific input states are the photon added squeezed vacuum state (PAS), the photon subtracted squeezed vacuum state (PSS) and squeezed number state (SNS).
The analysis in this paper hinges on our numerical results on the dependence of the teleportation fidelity on the squeeze parameter ($r$) for various values of the photon addition/subtraction number ($m$) in the case of specific input states.
We analyze our numerical results on teleportation in the light of various properties of resource states that have been considered in the literature to be crucial for QT, in particular, EPR correlation \cite{tp_yang,tp_lee,tp_wang,tp_agarwal,tp_zubairy} and squeezed vacuum affinity \cite{tp_illuminati}.

It was believed \cite{tp_yang,tp_agarwal,tp_zubairy} that EPR correlation could be a necessary/sufficient condition for QT. 
However, as argued in \cite{tp_lee,tp_wang}, EPR correlation is not always necessary for QT - a counterexample being that of the TMSV with photon added symmetrically in both modes that yields QT even without EPR correlation.
Furthermore for a large subset of states that we have considered in this paper, we have found that EPR correlation is not even sufficient for QT.
This result of ours in conjunction with the results of \cite{tp_lee,tp_wang} indicates that EPR correlation is \emph{neither necessary nor sufficient} for QT.
Although it has not been stated explicitly by Dell'Anno \textit{et. al.} \cite{tp_illuminati}, it is implicit in their work that squeezed vacuum affinity is a necessary ingredient for QT.
It so happens that for the states that they have considered, SVA is always nonzero.
Resource states for which SVA is non-zero zero may in principle yield QT. 
In fact, some of the resource states that we have considered in this paper do have this property.
However, we would like to emphasize here SVA is not non-zero, in general. It is easily seen that for any bipartite state other than that having the form $\langle n1,n1|\rho|m1,m1\rangle=\delta_{n1,n2}~\delta_{m1,m2}~ \langle n1,n2|\rho|m1,m2\rangle$, SVA vanishes.
In fact, in the case of most of the states we have considered, SVA becomes trivially zero.
Hence, it is clear that SVA can't be regarded as an essential ingredient for QT.

In view of the above discussion, the question arises as to what is the property of the resource states, besides entanglement, that contributes to QT when the resource states are not EPR correlated and SVA too is not non-zero.
In this paper we find that such a property is, in fact, the two-mode quadrature squeezing of the resource states as defined by Simon \emph{et. al.} \cite{qs_simon}.
Our numerical results on the class of non-Gaussian resource states studied in this paper indicate that two-mode quadrature squeezing is, indeed, a \emph{necessary condition} for QT, in the sense that in all cases where the resource state is not two-mode quadrature squeezed the fidelity of teleportation is $<1/2$, i.e., there is no QT. However, two-mode quadrature squeezing is not a sufficient condition.

The paper is organized as follows. 
In {\bf Sec. II} we present our numerical results on the teleportation of a coherent state with BS generated entangled non-Gaussian resource states. These states are obtained with various single mode non-Gaussian and nonclassical at one of the input ports while the other left with vacuum. 
In {\bf Sec III} we presents a detailed analysis of the entanglement, NG and SVA of the BS entangled states with a view to understand teleportation. 
In {\bf Sec. IV} we discuss EPR correlation of the resource states in light of the results on teleportation.
In {\bf Sec. V} we analyze our results in terms of two-mode quadrature squeezing character of the BS entangled resources. 
Here, we point out that two-mode quadrature squeezing is indeed a necessary condition for QT. 
{\bf Sec VI} contains summary of the work.

\section*{II. Teleportation of a Coherent State using the BS Generated non-Gaussian Entangled Resources}

In this section, we under take a qualitative study of QT with BS generated resource states.
For simplicity we consider nonclassical non-Gaussian single mode state at one of the input ports of the BS with vacuum at the other port.
The specific input states that we consider are the photon added squeezed vacuum state (PAS), the photon subtracted squeezed vacuum state (PSS) and the squeezed number state (SNS). 
These input states are mathematically described as,
\begin{subequations}
\label{psi_single}
\begin{align}
&|\psi_{\rm{pas}}\rangle =\frac{1}{\sqrt{\rm{N^{m}_{pas}}}} a^{\dagger m}S(r)|0\rangle \\
&|\psi_{\rm{pss}}\rangle =\frac{1}{\sqrt{\rm{N^{m}_{pss}}}} a^{m}S(r)|0\rangle \\
&|\psi_{\rm{sns}}\rangle =S(r)|m\rangle,
\end{align}
\end{subequations}
where, $S(r)=\exp[\frac{r}{2}(a^{\dagger 2}-a^{2})]$ is the single mode squeezing operator and the quantities $\rm{N^{m}_{pas}}$ and $\rm{N^{m}_{pss}}$ are defined by the relations $\rm{N^{m}_{pas}}=m!\mu^{m}P_{m}(\mu)$, $\rm{N^{m}_{pss}}=m!\nu^{2m}\sum_{k=0}^{m} \frac{m!}{(m-k)! k!}$ $(\frac{-\mu}{2\nu})^{k}\frac{H_{k}^{2}(0)}{k!}$, $\mu=\cosh r$ and $\nu=\sinh r$. 
Here $P_{n}(x)$ and $H_{n}(x)$ are respectively $n^{\rm{th}}$ order Legendre and Hermite polynomials.

A passive $50:50$ BS is described by the following transformation matrix between the input and the output mode operators,
\begin{equation}
\begin{pmatrix}
A \\
B
\end{pmatrix}=
\begin{pmatrix}
1/\sqrt{2} & 1/\sqrt{2}\\
-1/\sqrt{2} & 1/\sqrt{2}
\end{pmatrix}
\begin{pmatrix}
a \\
b
\end{pmatrix} .
\label{bs_trans}
\end{equation}
where $\lbrace A,B\rbrace$ and $\lbrace a,b\rbrace$ are the output and input mode operators respectively.
Since the input states are nonclassical, it is guaranteed that the corresponding BS output states will be entangled \cite{nc_bsent}.
It is well-known that entanglement is necessary for QT.
Next we analyze QT with these BS entangled resource states.

The teleportation protocol we consider is the standard Braunstein-Kimble (BK) \cite{TP_BK} protocol.
The performance/success of the teleportation is measured in terms of the fidelity of teleportation ($F$), defined as the overlap between the unknown input state and the output state (the retrieved state), $F=Tr[\rho_{\rm{in}}\rho_{\rm_{out}}]$. 
The evaluation $F$ becomes particularly simple in the characteristic function (CF) description \cite{TPF_CF}. 
The CF of an $n$ mode quantum optical state $\rho$ is defined as $\chi_{\rho}(\lbrace \lambda_{i} \rbrace)=Tr[\rho D(\lbrace \lambda_{i} \rbrace)]$ where $D(\lbrace \lambda_{i} \rbrace)=\Pi_{i=1}^{n}\exp [\lambda_{i} a^{\dagger}_{i}-\lambda^{*}_{i}a_{i}]$; $a_{i}$ being the $i^{\rm{th}}$ mode operator.
For any two-mode state $\rho_{\rm{AB}}$ as a resource, the fidelity of teleportation of an unknown input state $\rho_{\rm{in}}$ can be expressed as \cite{TPF_CF},
\begin{equation}
F=\int \frac{d^{2}\lambda}{\pi}~ \chi_{\rm{in}}(-\lambda)~\chi_{\rm{in}}(\lambda)~\chi_{\rm{AB}}(\lambda,\lambda^{*}),
\label{def_telfid}
\end{equation}
where, $\chi_{\rm{in}}(\lambda)$ and $\chi_{\rm{AB}}(\lambda,\lambda^{*})$ are the CFs of $\rho_{\rm{in}}$ and $\rho_{\rm{AB}}$ respectively. 
For simplicity we consider a coherent state as the unknown input state and BS generated entangled states as resource.
In this case, Eqn. (\ref{def_telfid}) simplifies to,
\begin{equation}
F=\int \frac{d^{2}\lambda}{\pi}~ e^{-\lambda^{2}}~\chi_{\rm{BS}}^{\rm{out}}(\lambda,\lambda^{*}) ,
\label{telfid_coh}
\end{equation}
where, $\chi_{\rm{BS}}^{\rm{out}}(\lambda,\lambda^{*})$ corresponds to the characteristic function of the BS output state.
Henceforth, we shall use Eq. \ref{telfid_coh} while discussing $F$. 
The maximum fidelity of teleportation of a coherent state attainable by a separable state in the BK protocol is $1/2$ \cite{TPF_CS}. 
Hence, $F> \frac{1}{2}$ indicates QT. 
\begin{figure}[h]
\includegraphics[scale=2.3]{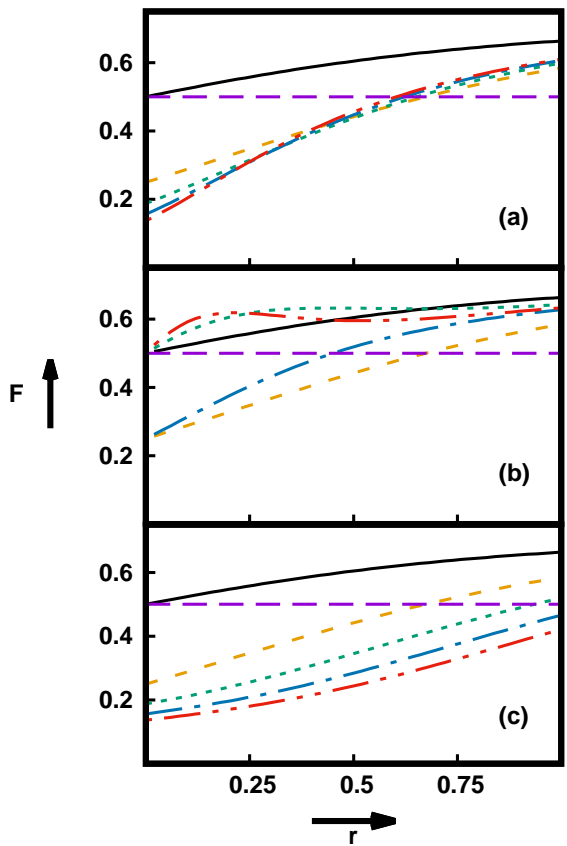}
\caption{(Color Online) Plot of $F$ vs $r$ for $m=0$ (black solid line), $1$ (yellow dashed line), $2$ (green dotted line), $3$ (blue dashed dotted line) and $4$ (red dashed double dotted line) with BS output states generated from the input states {\bf (a)} PAS, {\bf (b)} PSS and {\bf (c)} SNS. Violet long dashed line corresponds to the maximum limit for the "classical" teleportation, i.e, $1/2$. \label{fig_tf}}
\end{figure}

In Fig \ref{fig_tf} we plot the dependence of $F$ on the squeeze parameter $r$ and the number of photon addition/subtraction $m$ in the case of non-Gaussian BS output entangled states generated from single mode input states [Eq. (\ref{psi_single}a), (\ref{psi_single}b) and (\ref{psi_single}c)]. 
As is evident from Fig. (\ref{fig_tf}), in the case of all three input states, the teleportation fidelity $F$ exhibits a rather complex, in particular non-monotonic, dependence on the state parameters $r$ and $m$. 

The rest of the paper is devoted to understanding the various ramifications of the principle numerical results in Fig. \ref{fig_tf}. 
In the next few sections we shall assess the role of various attributes of the resource states, viz. entanglement, non-Gaussianity (NG), squeezed vacuum affinity (SVA) and EPR correlation on teleportation fidelity in respect of the results in Fig. \ref{fig_tf}.

\section*{III. Attributes of the Resource States I: Entanglement, NG and SVA}

In this section we essentially extend the analysis of Dell'Anno \emph{et. al.} \cite{tp_illuminati} to BS generated resource states.
It is pertinent to recall here the observation of Dell'Anno \emph{et. al.} \cite{tp_illuminati}, in the context of resource states generated by certain de-Gaussifications of the TMSV, that in order to achieve optimal teleportation, one has to tune values of entanglement, NG and SVA of the resource states.
The purpose of this section is to verify if this observation of Dell'Anno \emph{et. al.} is borne out in the case of BS generated resource states with $|\psi_{\rm{pas}}\rangle$, $|\psi_{\rm{pss}}\rangle$ and $|\psi_{\rm{sns}}\rangle$ at the input.

\subsection*{III-A. Entanglement and Teleportation Fidelity}

\begin{figure}[h]
\includegraphics[scale=2.3]{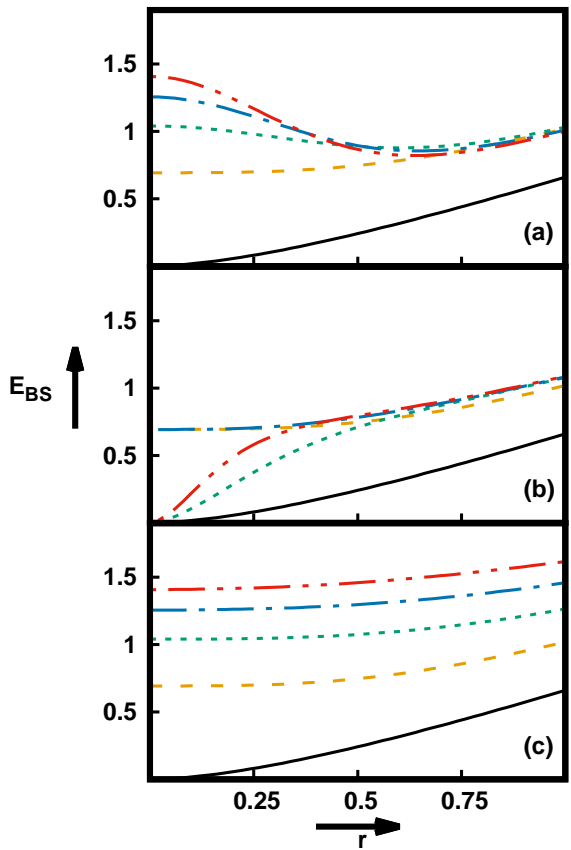}
\caption{(Color Online) Dependence of $E_{\rm{BS}}$ on $r$ for $m=0$ (black solid line), $1$ (yellow dashed line), $2$ (green dotted line), $3$ (blue dashed dotted line) and $4$ (red dashed double dotted line) for the input states {\bf (a)} PAS, {\bf (b)} PSS and {\bf (c)} SNS. \label{fig_ent}}
\end{figure}

We denote the BS generated entanglement with input $\vert\psi\rangle$ by $E_{\rm{BS}}^{|\psi\rangle}$. 
In Fig \ref{fig_ent}, we plot the dependence of BS entanglement for different input states. 
The specific dependence of $E_{\rm{BS}}^{|\psi_{\rm{pas}}\rangle}$ [Fig. \ref{fig_ent}(a)] and $E_{\rm{BS}}^{|\psi_{\rm{sns}}\rangle}$ [Fig. \ref{fig_ent}(c)] on $r$ and $m$ have already been discussed in detail in \cite{bose_kumar}. 
Here, we reproduce the figures for $E_{\rm{BS}}^{|\psi_{\rm{pas}}\rangle}$ and $E_{\rm{BS}}^{|\psi_{\rm{sns}}\rangle}$ from our previous work \cite{bose_kumar} for the sake of future discussion.
However, the results for $E_{\rm{BS}}^{|\psi_{\rm{pss}}\rangle}$ are new and we discuss them in some detail. 

In the case of $E_{\rm{BS}}^{|\psi_{\rm{pss}}\rangle}$ [Fig. \ref{fig_ent}(b)], we find that for small $r$ ($\leq 0.40$), odd photon subtracted states [$m=1,3$] are more entangled than the even photon subtracted states [$m=2,4$]. 
However, with increase in $r$, $E_{\rm{BS}}^{|\psi_{\rm{pss}}\rangle}$ for even photon subtracted states becomes higher than that for odd photon subtracted states. 
In general, $E_{\rm{BS}}^{|\psi_{\rm{pss}}\rangle}$, for all values of $m$, increases monotonically with increase in $r$. 

As it is quite explicit from Fig. \ref{fig_tf} and Fig. \ref{fig_ent}, the dependence of the fidelity of teleportation on input parameters $r$ and $m$ for different input states is very different from that of the respective BS output entanglement. 
In the cases of both PAS, PSS and SNS as input, BS output entanglement, for all non-zero values of $m$ and $r$, is always greater than that for the input Gaussian single mode squeezed vacuum state ($m=0$). 
However, in the case of teleporation, we observe that $F$ for all input states, except for the case of even PSS in the small $r$ ($\lesssim 0.60$) limit, is always smaller compared to the case of input squeezed vacuum state for all non-zero values of $m$ and $r$.

In the case of even PSS input, in the small $r$ ($\lesssim 0.30$) region, all input odd PSSs yields more entanglement at the output of BS than the input even PSSs.
However, $F$ for all even PSSs at BS input is greater than all input odd PSSs.
These results indicate the well-known fact that, although entanglement is necessary for QT, increase in entanglement does not always ensure increase in fidelity of teleportation.

\subsection*{III-B. NG and Teleportation Fidelity}

In this subsection we study how teleportation fidelity depends on the NG of the BS generated resource states.
There have been several proposals for quantification of the non-Gaussian character of any state in terms of Hilbert-Schmidt distance \cite{ngm_hsd}, relative entropy \cite{ngm_re}, Wehrl entropy \cite{ngm_we} \emph{etc}. In the present paper, we consider the Wehrl entropy based measure of NG. 

For a quantum state of light, described by the density operator $\rho$, its non-Gaussianity is defined as,
\begin{equation}
\delta(\rho)=H_{\rm{w}}(\rho^{\rm{G}})-H_{\rm{w}}(\rho),
\label{def_ng}
\end{equation} 
where $H_{\rm{w}}(\rho)$ [$=-\int\frac{d^{2}z}{\pi}Q_{\rho}(z)\log Q_{\rho}(z)$] is the Wehrl entropy of $\rho$ defined in terms of the Husimi-Kano $Q_{\rho}(z)$ [$=\langle z|\rho|z\rangle$] distribution.
Here $\rho^{\rm{G}}$ is the Gaussian counterpart of $\rho$, the state formed with the first and the second moments equal to those of $\rho$ itself. 

It is further shown by Ivan \emph{et. al.} \cite{ngm_we} that, in the case of product state input at any passive linear system like BS, NG of the output state becomes equal to the sum of NG of the input states, i.e.,
\begin{equation}
\delta(\rho_{\rm{out}})=\delta(\mathscr{U}_{\rm{BS}}(\rho_{a}\otimes \rho_{b})\mathscr{U}_{\rm{BS}}^{\dagger})=\delta(\rho_{a}) + \delta(\rho_{a}),
\label{NG_BsOut}
\end{equation}
where, $\mathscr{U}_{\rm{BS}}$ is the unitary operation corresponding to the evolution of the input state ($\rho_{a}\otimes \rho_{b}$) through BS.
In the current work we have considered the cases where one of the input ports BS is fed with single mode non-Gaussian states $\rho_{\rm{in}}$ while the other port is left with vacuum.
Since, vacuum ($|0\rangle$) is a Gaussian state with $\delta(|0\rangle)=0$, Eq. (\ref{NG_BsOut}) immediately implies that the NG of the BS generate resource states ($\rho_{\rm{out}}$) we have considered here is same as the NG of the corresponding input state $\rho_{\rm{in}}$. 
\begin{figure}[h]
\includegraphics[scale=2.1]{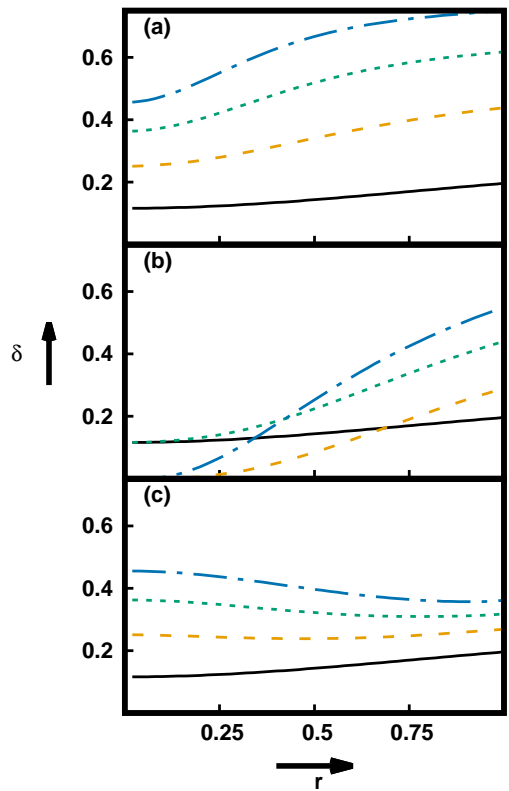}
\caption{(Color Online) Plot of $\delta$ of the BS output states vs $r$ for $m=1$ (black solid line), $2$ (yellow dashed line), $3$ (green dotted line) and $4$ (blue dashed dotted line) for the input states {\bf (a)} PAS, {\bf (b)} PSS and {\bf (c)} SNS. \label{fig_ng}}
\end{figure}

In Fig. \ref{fig_ng} we plot $\delta$ with $r$ for different $m$, for the BS output states generated from different input states. By $\delta^{|\psi\rangle}$ we denote the NG of the BS output state generated from the input state $|\psi\rangle$. 
It is clear from Fig. \ref{fig_tf} and Fig. \ref{fig_ng} that the fidelity of teleportation ($F$) does not depend monotonically on the NG ($\delta$) of the resource states. 
In the case of input PAS, $\delta$ increases monotonically with increase in both $m$ and $r$, while corresponding $F$ shows a non-monotonic dependence. 
In the case of input PSS, while the odd $m$ states are more non-Gaussian than the even $m$ states at low $r$ ($\lesssim 0.30$) limit, $F$ for input even PSSs is always higher than that for input odd PSSs.
Besides, in the case of input SNS, $\delta$ shows a non-monotonic dependence on $r$ for higher values of $m$ while corresponding $F$ is a monotonically increasing function of $r$ for all values of $m$.

\subsection*{III-C. SVA and Teleportation Fidelity}

Dell'Anno \emph{et. al.} \cite{tp_illuminati} identified yet another attribute called squeezed vacuum affinity ($\eta$) that entangled quantum optical resources must possess to achieve QT.
For any bipartite entangled state $\rho_{\rm{AB}}$, $\eta$ is defined as its maximal overlap with the TMSV ($|\xi(s)\rangle$),
\begin{equation}
\eta=\max_{s}|\langle \xi(s)|\rho|\xi(s)\rangle|^{2} .
\label{def_sva}
\end{equation} 

\begin{figure}[h]
\includegraphics[scale=2.5]{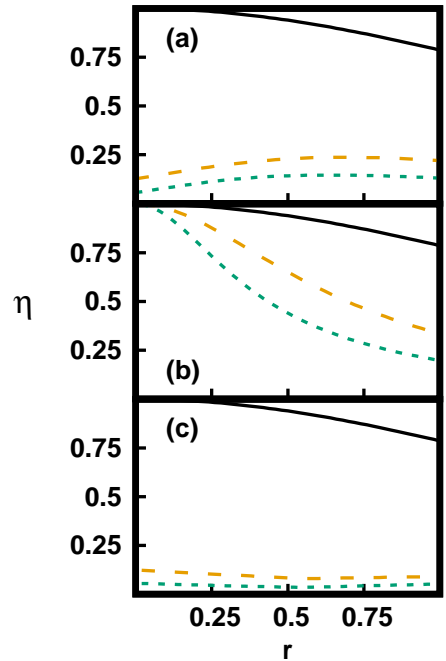}
\caption{$\eta$ for BS output states with input {\bf (a)} PAS, {\bf (b)} PSS and {\bf (c)} SNS for even $m$. We consider $m=0$ (black solid line), $2$ (yellow dashed line) and $4$ (green dotted line). \label{fig_sva}}
\end{figure}

First, We have analyzed the case of even photon added/subtracted states ($m=0,2,4$) at input of the BS for which the output states have nonzero $\eta$. 
In Fig. \ref{fig_sva}, we have shown the dependence of $\eta$ of the BS generated resource states for different input states.
As evident, in the case of all input states, $\eta$ for the BS output resource states decrease with increase in $r$ for different values of $m$. 
The maximum SVA is obtained for $r=0$ and $m=0$ that corresponds to the vacuum state ($|0\rangle$).

However, we have noticed that $\eta$ becomes trivially zero in the case of all input states with odd photon addition/subtraction.
This could be explained in the following way.
The state TMSV has a symmetric expansion in number state basis $|\xi(s)\rangle=\frac{1}{\mu_{s}}\sum_{k}\tau_{s}^{k}|k,k\rangle$, where $\mu_{s}=\cosh s$ and $\tau_{s}=\tanh s$. Let's now consider a bipartite state $\rho=\sum_{\substack{m,n\\ k,l}}~C_{m,n}^{k,l}$ $|m,n\rangle\langle k,l|$. The overlap between $|\xi(s)\rangle$ and $\rho$ is given by,
\begin{equation}
\rm{overlap}=\langle \xi(s)|\rho|\xi(s)\rangle= \frac{1}{\mu_{s}}\sum_{\substack{m,n\\ k,l}}~C_{m,n}^{k,l}~\tau_{s}^{m+k}~\delta_{m,n}~\delta_{k,l}.
\label{overlap}
\end{equation}

Evidently, in the case of a bipartite state $\rho$ for which the diagonal elements for all $m$ and $k$ vanish (e.g., $C_{m,m}^{k,k}=0$), SVA is identically zero. 
Note that a passive BS simply redistributes the photons in the input modes among the output modes. 
As a consequence, for all odd number ($m=2p+1$, $p$ is any positive integer) of photon added/subtracted states at input, BS output state have diagonal elements identically equal to zero, i.e, $C_{m,m}^{k,k}=0$ leading to $\eta=0$.

It is quite clear from Fig. \ref{fig_ent}, \ref{fig_ng} and \ref{fig_sva} for entanglement, NG and SVA respectively, that these attributes do not behave quite the same way as the teleprotation fidelity [Fig. \ref{fig_tf}], as far as their dependence on $r$ and $m$ is concerned.
In other words, $F$ depends non-monotonically on each of these attributes. 
One can't achieve a larger value of $F$ merely by increasing any one of these attributes.
Thus, our results in the case of those BS generated resource states for which SVA is non-zero are consistent with those of Dell'Anno \emph{et. al.} in the case of de-Gaussified two-mode squeezed vacuum states.

\section*{IV. Attributes of the Resource States II: EPR Correlation}

In recent years, besides entanglement, Einstein-Podolsky-Rosen (EPR) correlation \cite{epr_corr} of the two-mode resource states have been found to be an important ingredient in achieving QT \cite{tp_yang,tp_agarwal,tp_zubairy}.
However, Lee \emph{et. al.} \cite{tp_lee} and Wang \emph{et. al.} \cite{tp_wang} have have pointed to examples of states that yield QT even without EPR correlation.
In this section, we study this attribute in the case of BS generated non-Gaussian entangled resource states.

In the seminal paper on completeness of quantum mechanics \cite{EPR}, Einstein, Podolsky and Rosen proposed an ideal bipartite state which is a common eigenstate of the relative position and total momentum of the subsystems.
In the case of any two-mode quantum optical state one can define an EPR correlation parameter known as EPR uncertainty $\Delta_{\rm{EPR}}$ \cite{epr_corr} as 
\begin{align}
\Delta_{\rm{EPR}}&=\langle(\Delta(X_{\rm{A}}-X_{\rm{B}}))^{2}\rangle+\langle(\Delta(P_{\rm{A}}+P_{\rm{B}}))^{2}\rangle \nonumber \\
\begin{split}
&=2\big([1+\langle A^{\dagger}A\rangle+\langle B^{\dagger}B\rangle-\langle A^{\dagger}B^{\dagger}\rangle-\langle AB\rangle] \nonumber
\end{split}\\
&~~~~~~~~~~~~~~~~~~~~~~~~~[\langle A^{\dagger}\rangle-\langle B\rangle][\langle A\rangle-\langle B^{\dagger}\rangle]\big), \label{def_epr}
\end{align}
where, the quadrature operators $\lbrace X_{\rm{A}},P_{\rm{A}},X_{\rm{B}},P_{\rm{B}}\rbrace$ are defined as $X_{\rm{A}}=\frac{1}{\sqrt{2}}(A+A^{\dagger})$, $P_{\rm{A}}=\frac{1}{i\sqrt{2}}(A-A^{\dagger})$, $X_{\rm{B}}=\frac{1}{\sqrt{2}}(B+B^{\dagger})$ and $P_{\rm{B}}=\frac{1}{i\sqrt{2}}(B-B^{\dagger})$.
EPR uncertainty ($\Delta_{\rm{EPR}}$) being zero indicates perfect correlation between the modes.
The correlated state considered by Einstein \emph{et. al.} which is known as the EPR state \cite{epr_st}, could be realized in terms of TMSV in the limit of infinite squeezing strength ($r\rightarrow \infty$).
In the case of two-mode states with $\Delta_{\rm{EPR}}>0$, smaller the value of $\Delta_{\rm{EPR}}$ more correlated the modes are.
Further, as shown by Duan \emph{et. al.} \cite{epr_corr} $\Delta_{\rm{EPR}}<2$ indicates that the two-mode state is entangled.

In this section, we evaluate EPR correlation for the BS generated entangled resources for the different input non-Gaussian states we have considered in this paper. 
Using the transformation matrix for a $50$:$50$ BS [Eq. (\ref{bs_trans})], $\Delta_{\rm{EPR}}$ [Eq. (\ref{def_epr})] for the BS generated resource states can be expressed in terms of the input mode operators as,
\begin{align}
\Delta_{\rm{EPR}}&= 2\big( 1+\langle a^{\dagger}a\rangle + \langle b^{\dagger}b\rangle - \langle a^{\dagger}\rangle\langle a\rangle - \langle b^{\dagger}\rangle\langle b\rangle \big) - \nonumber \\
&~~~~~~~~~~~~~~~~~~ \big( \langle a^{\dagger 2}\rangle + \langle a^{2}\rangle -\langle a^{\dagger}\rangle^{2} - \langle a\rangle^{2} \big) - \nonumber \\
&~~~~~~~~~~~~~~~~~~ \big( \langle b^{\dagger 2}\rangle + \langle b^{2}\rangle -\langle b^{\dagger}\rangle^{2} - \langle b\rangle^{2} \big) . \label{def_epr_bs}
\end{align}

We have considered single mode nonclassical states at one of the input ports (say mode $a$) while other port (mode $b$) is left in the vacuum state. 
This leads to $\langle b\rangle=\langle b^{\dagger}\rangle=\langle b^{2}\rangle=\langle b^{\dagger 2}\rangle=\langle b^{\dagger}b\rangle=0$. 
Besides, for the input nonclassical states we have considered, $\langle a\rangle=\langle a^{\dagger}\rangle=0$ and $\langle a^{2}\rangle=\langle a^{\dagger 2}\rangle$. 
With these results, EPR uncertainty for the BS output states [Eq. \ref{def_epr_bs}] reduces to,
\begin{equation}
\Delta_{\rm{EPR}}=2\big( 1 + \langle a^{\dagger}a\rangle - \langle a^{2}\rangle \big) \label{def_epr_bs_in}.
\end{equation}

We denote the $\Delta_{\rm{EPR}}$ in the case of input state $|\psi\rangle$ as $\Delta_{\rm{EPR}}^{|\psi\rangle}$. Using the expression of Eq. \ref{def_epr_bs_in}, we find the analytic forms of the $\Delta_{\rm{EPR}}$, for input PAS, PSS and SNS as,
\begin{subequations}
\label{epr_psi_input}
\begin{align}
\Delta_{\rm{EPR}}^{|\psi_{\rm{pas}}\rangle}&=2\Big[ \frac{N^{m+1}_{\rm{pas}}}{N^{m}_{\rm{pas}}} + ~\frac{\mu^{2m}(m+2)!}{N^{m}_{\rm{pas}}}~\Big(\frac{\mu\nu}{2}\Big) \sum_{k=0}^{m} \begin{pmatrix}
m\\
k
\end{pmatrix} \nonumber\\
&~~~~~~~~~~~~~~~~~~~~~ \Big(\frac{-\nu}{2\mu}\Big)^{k}~\frac{H_{k}(0)H_{k+2}(0)}{(k+2)!} \Big], \\
\Delta_{\rm{EPR}}^{|\psi_{\rm{pss}}\rangle}&=2\Big[1+ \frac{N^{m+1}_{\rm{pss}}}{N^{m}_{\rm{pss}}} + ~\frac{\nu^{2m}(m+2)!}{N^{m}_{\rm{pss}}}~\Big(\frac{\mu\nu}{2}\Big)~ \sum_{k=0}^{m} \begin{pmatrix}
m\\
k
\end{pmatrix} \nonumber \\
&~~~~~~~~~~~~~~~~~~~~~~ \Big(\frac{-\mu}{2\nu}\Big)^{k}~ \frac{H_{k}(0)H_{k+2}(0)}{(k+2)!} \Big], \\
\Delta_{\rm{EPR}}^{|\psi_{\rm{sns}}\rangle}&=2\Big[1+ m(\mu-\nu)^{2}-\nu(\mu-\nu) \Big],
\end{align}
\end{subequations}
where, $\mu=\cosh r$, $\nu=\sinh r$ and $H_{n}(x)$ is the $n^{\rm{th}}$ order Hermite polynomial. The expression $\begin{pmatrix}
m\\
k
\end{pmatrix}$ is the binomial coefficient and the normalization constants $N^{m}_{\rm{pas}}$ and $N^{m}_{\rm{pss}}$ are defined in Eq. (\ref{psi_single}).

\begin{figure}[h]
\includegraphics[scale=2.2]{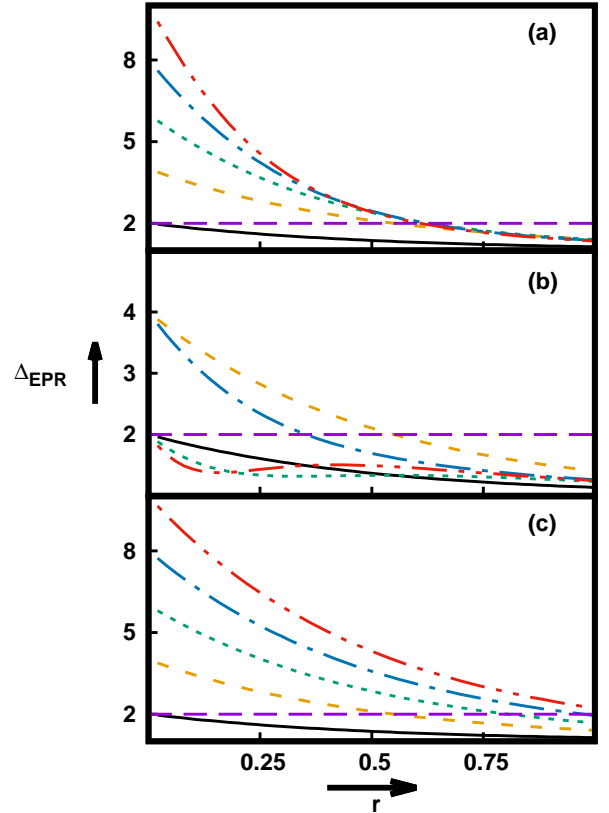}
\caption{(Color Online) Dependence of $\Delta_{\rm{EPR}}$ on $r$ for different $m=0$ (black solid line), $1$ (yellow dashed line), $2$ (green dotted line), $3$ (blue dashed dotted line) and $4$ (red dashed double dotted line) for input {\bf (a)} PAS, {\bf (b)} PSS and {\bf (c)} SNS. The long dashed violet line correspoonds to $\Delta_{\rm{EPR}}=2.0$. \label{fig_epr}}
\end{figure}

In Fig. \ref{fig_epr} we have plotted $\Delta_{\rm{EPR}}$ as a function of $r$ for various values of $m$ for the BS output states, generated from the input single mode states.
It is evident from a comparison of Fig. \ref{fig_tf} for the teleportation fidelity and Fig. \ref{fig_epr} for the EPR correlation in the case of BS generated resource states with either of the $|\psi_{\rm{pas}}\rangle$, $|\psi_{\rm{pss}}\rangle$ and $|\psi_{\rm{sns}}\rangle$, that there exists particular region of $r$ where resource states are EPR correlated ($\Delta_{\rm{EPR}}<2$) yet they don't yield QT ($F>1/2$). 

This leads to the conclusion that EPR correlation is not sufficient for QT.
Further, as shown by Lee \emph{et. al.} \cite{tp_lee} and Wang \emph{et. al.} \cite{tp_wang} in the case of non-Gaussian entangled states, QT can be achieved even when the resource state is not EPR correlated, i.e. $\Delta_{\rm{EPR}}>2$.
Thus, in view of our results together with the results of \cite{tp_lee,tp_wang}, we conclude that EPR correlation is \emph{neither necessary nor sufficient} for QT.

Let us summarize our analysis of the various attributes of the resource states so far.
The results of Sec. II-C lead us to conclude that SVA, as it is not non-zero in general and in particular in the case of BS generated resource states, it cannot be regarded.
Further, when it is non-zero it is not even sufficient.
Further, the results of Sec. IV make it clear that EPR correlation is neither necessary nor sufficient.
In the backdrop of these results the question of what other attributes of the resource states, beside entanglement, play an essential role in QT remains open.

We propose yet another attribute of resource states that has not been considered in the literature in the context of QT, namely the $U(2)$-invariant two-mode quadrature squeezing as defined by Simon \emph{et. al.} \cite{qs_simon}.
In the next section we examine the role of this attribute in the context of QT.

\section*{V. Attributes of the Resource States III: Two-mode Quadrature Squeezing}

We recall here the definition of the $U(2)$-invariant two-mode quadrature squeezing as defined by Simon \emph{et. al.} \cite{qs_simon}.
Let's consider an two-mode quantum state of light $\rho$ with mode annihilation operators $a_{k}$ [$k=1,2$] satisfying the commutation relations,
\begin{align}
[a_{k},a^{\dagger}_{l}]&=\delta_{k,l}~~\rm{and} \nonumber \\
[a_{k},a_{l}]&=[a_{k}^{\dagger},a_{l}^{\dagger}]=0 .
\end{align}

In terms of the quadrature components, namely $x_{k}=\frac{1}{\sqrt{2}}(a_{k}+a^{\dagger}_{k})$ and $p_{k}=\frac{1}{i\sqrt{2}}(a_{k}-a^{\dagger}_{k})$ ($k=1,2$), one can define a column vector as $\overrightarrow{R}=(x_{1},p_{1},x_{2},p_{2})^{T}$, where "$T$" stands for transposition.
The variance matrix of a two-mode state $\rho$ can be written in a compact form as $V_{k,l}=\frac{1}{2}~\rm{Tr}[\rho\lbrace \Delta R_{k},\Delta R_{l}\rbrace]$, where $\Delta R_{k}=R_{k}-\rm{Tr}[\rho R_{k}]$.
The state $\rho$ is said to quadrature squeezed if
\begin{equation}
\lambda_{\rm{min}}<\frac{1}{2},
\label{def_sq}
\end{equation}
where, $\lambda_{\rm{min}}$ is the least eigenvalue of its variance matrix $V$ \cite{qs_simon}. 
Accordingly, the degree of quadrature squeezing is defined as,
\begin{equation}
f_{\rm{sq}}=\frac{1}{\sqrt{2\lambda_{\rm{min}}}} ,
\label{def_degsq}
\end{equation}
and in line with Eq. \ref{def_sq}, the state is said to be quadrature squeezed if $f_{\rm{sq}}>1$.
Henceforth, throughout the rest of the paper, any discussion on two-mode quadrature squeezing will correspond to the $U(2)$-invariant squeezing as described by Eq. (\ref{def_sq}) and Eq. (\ref{def_degsq}) accordingly.

In the following, we shall compute two-mode quadrature squeezing for two different class of entangled resources, namely, {\bf (a)}: states obtained by symmetrically single photon addition/subtraction on TMSV and two-mode squeezed number states considered by Dell'Anno {\em et. al.} \cite{tp_illuminati} and {\bf (b)}: BS generated entangled states considered in this paper.

\subsection*{V-A. Two-mode quadrature squeezing for states considered by Dell'Anno {\em et. al.}}

Let's denote the states considered in \cite{tp_illuminati} by, 
\begin{subequations}
\label{psi_illuminati}
\begin{align}
|\psi_{\rm{TMSV}}\rangle&=S_{a,b}(r)|0,0\rangle, \\
|\psi_{\rm{tmpa}}\rangle&=\frac{1}{N_{+}}a^{\dagger}b^{\dagger}S_{a,b}(r)|0,0\rangle, \\
|\psi_{\rm{tmps}}\rangle&=\frac{1}{N_{-}}abS_{a,b}(r)|0,0\rangle, \\
|\psi_{\rm{tmsn}}\rangle&=S_{a,b}(r)|1,1\rangle,
\end{align}
\end{subequations}
where, $S_{a,b}(r)=e^{r(a^{\dagger}b^{\dagger}-ab)}$, $N_{+}$ and $N_{-}$ are the normalization constants.

\begin{figure}[h]
\hspace*{-1 cm}
\includegraphics[scale=2.2]{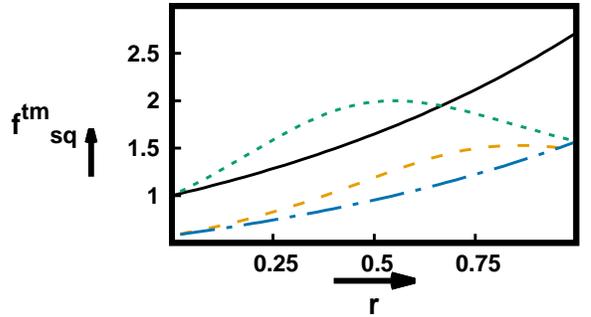}
\caption{(Color Online) Plot of $f_{\rm{sq}}^{\rm{tm}}$ vs $r$ for TMSV (black solid line), $|\psi_{\rm{pa}}\rangle$ (yellow dashed line), $|\psi_{\rm{ps}}\rangle$ (green dotted line) and $|\psi_{\rm{sn}}\rangle$ (blue dashed dotted line). \label{fig_tmsq_illu}}
\end{figure}

We obtain analytic expressions for the $\lambda_{\rm{min}}$ for the states [Eq. (\ref{psi_illuminati}a), (\ref{psi_illuminati}b), (\ref{psi_illuminati}c) and (\ref{psi_illuminati}d)] as,
\begin{align}
\lambda_{\rm{min}}\big( |\psi_{\rm{TMSV}}\rangle \big)&=\frac{1}{2}-\nu(\mu-\nu), \nonumber \\
\lambda_{\rm{min}}\big( |\psi_{\rm{tmpa}}\rangle \big)&=\frac{1}{2}+(1-\tau)(1-3\tau+\tau^{2}-\tau^{3}) , \nonumber \\
\lambda_{\rm{min}}\big( |\psi_{\rm{tmps}}\rangle \big)&=\frac{1}{2}-2\tau(1-\tau)(1-\tau+\tau^{2}) , \nonumber \\
\lambda_{\rm{min}}\big( |\psi_{\rm{tmsn}}\rangle \big)&=\frac{1}{2}+(\mu-2\nu)(\mu-\nu),
\end{align}
where, $\mu=\cosh r$, $\nu=\sinh r$ and $\tau=\tanh r$. 
The degree of squeezing for the states, then, is calculated using Eq. (\ref{def_degsq}). 
In Fig. \ref{fig_tmsq_illu}, we plot the dependence of the degree of two-mode quadrature squeezing ($f_{\rm{sq}}^{\rm{tm}}$), for states given in Eq. (\ref{psi_illuminati}a), (\ref{psi_illuminati}b) and (\ref{psi_illuminati}c), upon squeezing strength $r$. 
We also plot, in the same figure, $f_{\rm{sq}}^{\rm{tm}}$ for TMSV as reference.

The degree of squeezing ($f_{\rm{sq}}^{\rm{tm}}$) for TMSV is found to be always greater than unity for all non-zero values of $r$. With increase in $r$, $f_{\rm{sq}}^{\rm{tm}}$ increases monotonically.
In the case of $|\psi_{\rm{pa}}\rangle$, we notice that the state shows two-mode squeezing ($f_{\rm{sq}}^{\rm{tm}}>1.0$) beyond $r\sim 0.30$. 
However, it leads to quantum teleportation for higher $r$ values \cite{tp_illuminati}.
In the case of $|\psi_{\rm{tmps}}\rangle$, we observe the presence of two-mode squeezing for all values of $r$ which falls in line with the curve for corresponding teleportation fidelity \cite{tp_illuminati}. 
It is worth noting that for a small squeeze parameter ($r\lesssim 0.65$), photon subtracted TMSV is more two-mode quadrature squeezed than the TMSV; however, for higher $r$ ($\gtrsim 0.70$) TMSV becomes more squeezed. 
In comparison with the specific curve for fidelity of teleportation \cite{tp_illuminati}, two-mode quadrature squeezing appears to be necessary for QT; however, not sufficient. 
In the case of $|\psi_{\rm{tmsn}}\rangle$ we find the dependence of $f_{\rm{sq}}^{\rm{tm}}$ on $r$ very similar to the case of $|\psi_{\rm{tmpa}}\rangle$.

\subsection*{V-B. Two-mode quadrature squeezing for the BS generated entangled states with specific input states}

Using the relation between variance matrices of the input and output state of a BS, it is easy to show (Appendix A) that the $\lambda_{\rm{min}}$ for the BS output states is given by, $\lambda_{\rm{min}}=\min [1/2,\Delta Q]$, where, $\Delta Q$ is the value of the uncertainty of the squeezed quadrature of the input state. We denote the the degree of squeezing for the BS output states as $f_{\rm{sq}}^{\rm{bs}}$.

\begin{figure}[h]
\hspace*{-1 cm}
\includegraphics[scale=2.0]{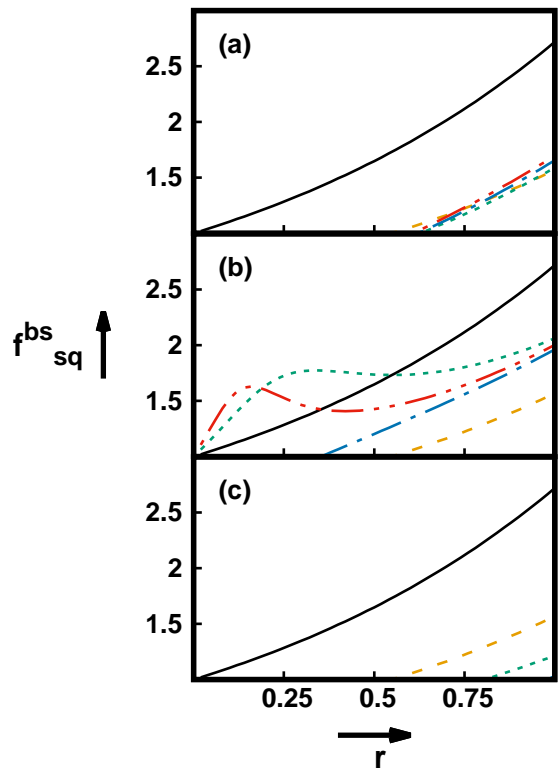}
\caption{(Color Online) Plot of $f_{\rm{sq}}^{\rm{bs}}$ vs $r$ for different $m=0$ (black solid line), $1$ (yellow dashed line), $2$ (green dotted line), $3$ (blue dashed dotted line) and $4$ (red dashed double dotted line) for the BS output states with single mode {\bf (a)} PAS, {\bf (b)} PSS and {\bf (c)} SNS at input. \label{fig_tmsq_bs}}
\end{figure}

In Fig. \ref{fig_tmsq_bs} we show the dependence of $f_{\rm{sq}}^{\rm{bs}}$ on $r$ for the BS output two-mode states generated from input PAS, PSS and SNS. 
In the case of input PAS [Fig. \ref{fig_tmsq_bs}(a)], $f_{\rm{sq}}^{\rm{bs}}$, for all $m\geq 1$, becomes greater than unity beyond a moderate squeezing strength ($r\gtrsim 0.60$). 
However, these states yield QT ($F>1/2$) for higher $r$.
In comparison to the results on $F$ [Fig. \ref{fig_tf}(a)], it explains the absence of QT below $r\sim 0.60$. 

In the case of input PSS, all the even photon subtracted states [$m=2$, $4$] as well as no photon subtracted state [$m=0$] possess two-mode quadrature squeezing ($f_{\rm{sq}}^{\rm{bs}}>1.0$) [Fig. \ref{fig_tmsq_bs}(b)] for all values of $r$. However, all the odd photon subtracted states attain $f_{\rm{sq}}^{\rm{bs}}>1.0$ for higher values of $r$.  
In comparison to the corresponding results on $F$ [Fig. \ref{fig_tf}(b)], it is clear that the states, we consider here, yield quantum teleportation provided they possess two-mode quadrature squeezing.

In the case of input SNS, we observe that $f_{\rm{sq}}^{\rm{bs}}$ [Fig. \ref{fig_tmsq_bs}(c)] for $m\neq 0$ becomes greater than unity for high values of $r$. The threshold value of $r$ for two-mode squeezing ($f_{\rm{sq}}^{\rm{bs}}>1.0$) increases with the increase in $m$. In the case of corresponding results on $F$ [Fig. \ref{fig_tf}(c)] also, we notice that for $m\neq 0$ states quantum teleportation ($F>1/2$) is attained for higher values of $r$.

It is noteworthy that all the BS output resource states that we have considered attain two-mode quadrature squeezing, depending upon the value of $m$, beyond a certain value of squeeze parameter $r$.
This could be explained in the following manner.
Using the relation between the variance of matrix of the state at input of the BS and that of the output state, it can be easily shown (Appendix) that the output state is quadrature squeezed if and only if the input single mode is quadrature squeezed.
Since, the input single mode states become quadrature squeezed ($f_{\rm{sq}}>1$) beyond a moderate value of squeeze parameter $r$, depending upon the value of $m$, the same is reflected in the quadrature squeezing of the output states.

A close examination of the numerical results on $f_{\rm{sq}}^{\rm{tm}}$ for the states considered by Dell'Anno {\em et. al.} \cite{tp_illuminati} as well as the BS generated states that we have considered in this work, indicates that {\em two-mode quadrature squeezing is necessary for QT}. However, two-mode quadrature squeezing is not a sufficient condition. 

In this connection it is instructive to examine if two-mode quadrature squeezing is necessary for QT in the case of Gaussian resource states.
In fact, it turns out (as we show in the next subsection) that in the case of symmetric Gaussian states two-mode quadrature squeezing is indeed necessary for QT.

\subsection*{V-C. Quantum Teleportation with symmetric Gaussian resource states}

Let's consider a symmetric Gaussian state with two-mode variance matrix $V$ of the following specific form, 
\begin{equation}
V=\begin{bmatrix}
\eta& 0& c& 0\\
0& \eta& 0& -c\\
c& 0& \eta& 0\\
0& -c& 0& \eta
\end{bmatrix} .
\label{vm_gen}
\end{equation} 

The necessary condition on $V$ (set by the uncertainty relation) to be a bona fide quantum variance matrix is that its symplectic eigenvalues ($\kappa_{i}$, $i=1,2$) (elements in the Williamson's diagonal form) must be no less than $1/2$, i.e. $\kappa_{i}\geq 1/2$. 
These, symplectic eigenvalues are obtained as the ordinary eigenvalues of $|iV\Omega|$, where,
\begin{equation}
\Omega=\begin{bmatrix}
J& 0\\
0& J
\end{bmatrix}~;~J=\begin{bmatrix}
0& 1\\
-1& 0
\end{bmatrix} .
\label{def_symplectic_metric}
\end{equation}

The condition $\kappa_{i}\geq 1/2$, for the variance matrix $V$ [Eq. \ref{vm_gen}] leads to,
\begin{equation}
\sqrt{(\eta+c)(\eta-c)}\geq 1/2
\label{cond_vm_bf} .
\end{equation}

According to the condition of two-mode quadrature squeezing as defined by Simon {\it et. al.} \cite{qs_simon}, the variance matrix $V$ is said to be quadrature squeezed if its "{\it least eigenvalue}" becomes less than $1/2$. For the variance matrix $V$ given in Eq. \ref{vm_gen}, its eigenvalues are $l=\eta\pm c$. Evidently, the condition of two-mode quadrature squeezing for $V$ yields
\begin{equation}
l_{\rm{min}}=\eta-c<1/2 .
\label{cond_vm_qs}
\end{equation}

Let's now look at the teleportation of the coherent state with the Gaussian resource states. For any Gaussian state with variance matrix $V=\begin{bmatrix}
A& C\\
C^{T}& B
\end{bmatrix}$, where, $A$, $C$ and $B$ are $2\times 2$ matrices, the fidelity of teleportation of a coherent state (Eq. \ref{def_telfid}) becomes \cite{Pirandola_LasPhys},
\begin{equation}
F=\frac{1}{\sqrt{\det[\mathscr{M}]}},
\label{def_tf_vm}
\end{equation}
where, $\mathscr{M}=A-\lbrace \sigma_{z},C\rbrace+\sigma_{z}B\sigma_{z}+I$. $\sigma_{z}$ is the Pauli spin matrix, $\sigma_{z}=\begin{bmatrix}
1& 0\\
0& -1\\
\end{bmatrix}$.

For the symmetric Gaussian states with variance matrix given in Eq. \ref{vm_gen}, we have $B=A=diag(\eta,\eta)$ and $C=C^{\rm{T}}=diag(c,-c)$. This leads to $\mathscr{M}=diag(1+2\overline{\eta-c},1+2\overline{\eta-c})$ with $\det [\mathscr{M}]=(1+2\overline{\eta -c})^{2}$. Now the condition of QT, i.e., $F>1/2$, leads to,
\begin{equation}
\sqrt{\det[\mathscr{M}]}\leq 2 \Rightarrow \eta-c \leq 1/2 .
\label{cond_vm_qt}
\end{equation}

Evidently,
the condition for quantum teleportation
(Eq. \ref{cond_vm_qt}) and
and the condition for quadrature squeezing (Eq.  \ref{cond_vm_qs}),
are identical. This implies that quadrature squeezing is a necessary
condition for QT with symmetric Gaussian resouce states. Further it
also implies that, in this case, it is also sufficient. However note
that we have considered teleportation of a coherent state via
Braunstein-Kimble protocol. We do
not expect that two-mode quadrature squeezing would be sufficient
in the case of teleportation of a general single-mode state and with general
Gaussian resource states.

In view of the result for the symmetric Gaussian states obtained above, that two-mode quadrature squeezing is necessary for QT, it is plausible that in the case of non-Gaussian entangled resource states as well {\it two-mode quadrature squeezing is necessary for QT}. 
 
\section*{VI. Conclusion}

In summary, we have studied QT with a class of non-Gaussian resource states.
These resource states are generated by a passive BS with specific single mode non-Gaussian states at one of the input ports, viz., the photon added squeezed vacuum state, the photon subtracted squeezed vacuum state and squeezed number state.
In contrast, the non-Gaussian resource states studied in the literature in the context of QT normally are those generated from various de-Gaussifications of the TMSV.
The analysis in this paper hinges on our numerical results on the dependence of the teleportation fidelity on the squeeze parameter ($r$) for various values of the photon addition/subtraction ($m$), in the case of different resource states.

Firstly, we have extended the analysis of Dell'Anno \emph{et. al.} \cite{tp_illuminati} to the BS generated non-Gaussian resource states and studied in detail the dependence of QT on entanglement, NG and SVA.
While Dell'Anno \emph{et. al.} used the Hilbert-Schmidt distance based NG measure, we instead have used the Wehrl entropy based measure.
Consistent with the results of Dell'Anno \emph{et. al.}, we have found that the teleportation fidelity doesn't depend monotonically on either of these properties but one has to tune the values of these to achieve optimal QT fidelity.

Our next focus has been to identify what all attributes of the resource states, apart from entanglement, are necessary and/or sufficient for QT.
To this end, we have studied SVA and EPR correlation which have been considered in the literature as being critical for QT.
However, we have found that SVA is, in general, not non-zero for all resource states.
In particular, it turns out to be zero in most of the cases for the class of states that we have considered.
On the other hand, while the fact that EPR correlation is not necessary for QT has been known in the literature \cite{tp_wang,tp_lee}, numerical results on our class of states indicate that it is not also sufficient.

We have proposed that two-mode $U(2)$-invariant squeezing \cite{qs_simon} is an appropriate attribute to consider in this context.
Our numerical results on both the class of BS generated non-Gaussian resource states as well as other de-Gaussified TMSV lead us to the conclusion that $U(2)$-invariant squeezing is, in fact, a \emph{necessary condition} that all resource states must satisfy.
To argue that this is a plausible conclusion we have given an analytical proof, in the case of symmetric Gaussian states, that $U(2)$-invariant two-mode quadrature squeezing is indeed \emph{necessary} for QT.
It turns out that in the special case of QT of a coherent state via Braunstein-Kimble protocol, and with symmetric Gaussian resouce states, two-mode quadrature squeezing is also sufficient. 
It would be nice to give an analytical proof that two-mode quadrature squeezing is necessary for all Gaussian entangled resource states.
We shall return to this question elsewhere.

\section*{Appendix: Least Eigenvalue of the Variance Matrix of BS Output States Generated from Single Mode Input States}

Here, we discuss the least eigenvalue of the two-mode variance matrix of the BS output states generated from a single mode nonclassical state at one of the input ports while the other port is left with vacuum. Let's consider the column vectors $R_{\rm{in}}$ and $R_{\rm{out}}$ for the input and output quadrature operators as
\begin{equation}
R_{\rm{in}}=\begin{bmatrix}
x_{a} \\
p_{a} \\
x_{b} \\
p_{b} \\
\end{bmatrix},~~
R_{\rm{out}}=\begin{bmatrix}
x_{A} \\
p_{A} \\
x_{B} \\
p_{B} \\
\end{bmatrix} .
\end{equation} 
The quadrature operators $x_{i}$, $p_{i}$ corresponding to annihilation and creation
operators $a_{i}, a^{\dagger}_{i}$ are defined as $x_{i}=\frac{1}{\sqrt{2}}(a_{i}+a^{\dagger}_{i})$ and 
and $p_{i}=\frac{1}{i\sqrt{2}}(a_{i}-a^{\dagger}_{i})$.

Using the transformation matrix between input and output mode operators Eq.
(\ref{bs_trans}) for a $50:50$ BS, it is easy to show that $R_{\rm{out}}$
is related to $R_{\rm{in}}$ by the transformation, $R_{\rm{out}}=S R_{\rm{in}}$,
i.e.,
\begin{equation}
\begin{bmatrix}
x_{A} \\
p_{A} \\
x_{B} \\
p_{B} \\
\end{bmatrix}=\begin{bmatrix}
\frac{1}{\sqrt{2}}& 0& \frac{1}{\sqrt{2}}& 0\\
0& \frac{1}{\sqrt{2}}& 0& \frac{1}{\sqrt{2}}\\
-\frac{1}{\sqrt{2}}& 0& \frac{1}{\sqrt{2}}& 0\\
0& -\frac{1}{\sqrt{2}}& 0& \frac{1}{\sqrt{2}}
\end{bmatrix}
\begin{bmatrix}
x_{a} \\
p_{a} \\
x_{b} \\
p_{b} \\
\end{bmatrix} .
\label{tran_bs_coor}
\end{equation}
It is well known that, under the linear transformation $S$, the input variance matrix $V_{\rm{in}}$ transforms as $SV_{\rm{in}}S^{T}$. In this paper we consider the class of states for which the input variance matrix $V_{\rm{in}}$ is given by,
\begin{equation}
V_{\rm{in}}=\begin{bmatrix}
\eta_{a}& 0& 0& 0\\
0& \zeta_{a}& 0& 0\\
0& 0& \frac{1}{2}& 0\\
0& 0& 0& \frac{1}{2}
\end{bmatrix} .
\end{equation}

Using the transformation $S$ given in Eq. (\ref{tran_bs_coor}) we get the output variance matrix as,
\begin{equation}
V_{\rm{out}}=SV_{\rm{in}}S^{T}=\begin{bmatrix}
\frac{\eta_{a}+1/2}{2}& 0& \frac{-\eta_{a}+1/2}{2}& 0\\
0& \frac{\zeta_{a}+1/2}{2}& 0& \frac{-\zeta_{a}+1/2}{2}\\
\frac{-\eta_{a}+1/2}{2}& 0& \frac{\eta_{a}+1/2}{2}& 0\\
0& \frac{-\zeta_{a}+1/2}{2}& 0& \frac{\zeta_{a}+1/2}{2}
\end{bmatrix} .
\end{equation}

It can be easily shown that the least eigenvalue of $V_{\rm{out}}$ is given by $\lambda_{\rm{min}}=\min[1/2,\eta_{a},\zeta_{a}]$. To be specific, let's assume
$\eta_{a}\geq\zeta_{a}$. In this case the minimum eigenvalue will be given
by $\lambda_{\rm{min}}=\min[1/2,\zeta_{a}]$.

\end{document}